\documentclass[twocolumn,aps,prr,showpacs,longbibliography,superscriptaddress]{revtex4-1}
\usepackage{amssymb,latexsym}
\usepackage{mathrsfs}
\usepackage{lipsum}
\usepackage{graphicx}
\usepackage[caption=false]{subfig}
\usepackage{mathtools}
\usepackage{epstopdf}
\usepackage{xcolor}
\usepackage[colorlinks, linkcolor=red, anchorcolor=green, citecolor=green, urlcolor=blue]{hyperref}
\DeclareGraphicsExtensions{.pdf,.jpg,.png,.eps}
\usepackage[toc,page,title,titletoc,header]{appendix}
\usepackage{etoolbox}
\usepackage{breqn}

\makeatletter
\let\cat@comma@active\@empty
\makeatother

\begin{document}
	
\title{Loss-tolerant simultaneous two-parameter estimation in a two-channel optical network}

\author{Jian-Dong Zhang}
\email[]{zhangjiandong@jsut.edu.cn}
\affiliation{School of Mathematics and Physics, Jiangsu University of Technology, Changzhou 213001, China}
\affiliation{Jiangsu Key Laboratory of Quantum Information Science and Technology, Nanjing University, China}
\author{Shuai Wang}
\affiliation{School of Mathematics and Physics, Jiangsu University of Technology, Changzhou 213001, China}
\affiliation{Jiangsu Key Laboratory of Quantum Information Science and Technology, Nanjing University, China}
\author{Lili Hou}
\affiliation{School of Mathematics and Physics, Jiangsu University of Technology, Changzhou 213001, China}
\affiliation{Jiangsu Key Laboratory of Quantum Information Science and Technology, Nanjing University, China}
\author{Xuechun Zhao}
\affiliation{School of Mathematics and Physics, Jiangsu University of Technology, Changzhou 213001, China}
\affiliation{Jiangsu Key Laboratory of Quantum Information Science and Technology, Nanjing University, China}

\date{\today}
	
\begin{abstract}
Quantum multi-parameter estimation, which exploits quantum resources such as entanglement and squeezing, is revolutionizing performance across diverse applications, including sensing and imaging.
In this work, we propose a protocol for simultaneous estimation of a phase shift and a transmittance parameter in a two-channel optical network.
We employ a Schr\"odinger's cat state along
with a coherent state as the probe, and projective measurement is utilized as the readout.
A comparison of the classical and quantum Fisher information matrices indicates that our protocol is approximately optimal.
The estimation sensitivity of the phase shift and that of the transmittance parameter reach the Heisenberg limit and the sub-shot-noise limit, respectively.
In addition, our protocol exhibits excellent tolerance to photon loss. 
Our findings may promote quantum multi-parameter estimation and its development in related fields.
\end{abstract}

\maketitle

\section{Introduction}

Measurements are an important physical process that can be utilized to acquire information about physical quantities.
Precision measurements, with the main goal of improving measurement performance and expanding measurement capabilities, serve as a crucial cornerstone in modern science.
Currently, the performance of classical measurement techniques has reached a bottleneck.
Quantum precision measurements are an important technology that breaks through the traditional limit of estimating unknown parameters.
By harnessing quantum effects such as superposition, entanglement, and coherence, the measurement performance can be improved from the shot-noise limit scaling of $1/\sqrt{N}$ to the Heisenberg limit scaling of $1/N$, where $N$ is the number of particles involved in the measurements. 
Quantum precision measurements can not only achieve a significant improvement in measurement performance, but also enable the completion of many tasks that classical technologies are incapable of handling, thus boasting broad application prospects.

Phase estimation, as a standard approach for sensing the variation of physical quantities, is at the core of the field of quantum precision measurements. 
During the past decades, phase estimation based on quantum states and optical interferometers has been the focus of related studies.
For linear interferometers, it has been proven that many probe states can achieve Heisenberg-limited phase sensitivity, including Gaussian states such as coherent and squeezed vacuum states \cite{PhysRevLett.100.073601,Schafermeier:18,PhysRevA.81.033819,PhysRevA.94.023834,zhong2020quantum}, two-mode squeezed vacuum states \cite{PhysRevLett.104.103602,PhysRevLett.130.070801}, and non-Gaussian states such as N00N states \cite{07510802091298,PhysRevLett.112.103604,PhysRevLett.130.123603} and entangled coherent states \cite{PhysRevLett.111.173601,PhysRevLett.107.083601,PhysRevA.88.043832,PhysRevA.101.012332}. 
For nonlinear interferometers, the achievement of phase sensitivity at the Heisenberg limit has also been extensively discussed. 
In addition, considerable efforts have been made to the subject of transmittance estimation \cite{PhysRevA.79.040305,PhysRevA.84.022334,Dowran_2022,Jonsson_2022,PhysRevA.102.052603,radia2026}.
This is closely related to the sensing of target reflectivity, sample absorption, and channel loss.
These studies can be used for widespread quantum information tasks, including quantum illumination, sensitive radar systems, and quantum noise sensing.

In recent years, with the increasing complexity of practical application scenarios, multi-parameter estimation has gained increasing interest in the community of quantum precision measurements. 
On the one hand, the estimation of multiple parameters belonging to the same degree of freedom has been widely analyzed, such as the estimation of multiple phase shifts, displacement parameters, and spatial separation parameters. \cite{PhysRevA.105.012607,PhysRevA.94.062312,ciampini2016quantum,PhysRevA.100.032104,PhysRevA.111.062408,qg9z-bn5l}. 
On the other hand, there has been a lot of interests in the estimation of multiple parameters belonging to different degrees of freedom, phase shift and loss parameter of an interferometer \cite{PhysRevA.89.023845,PhysRevLett.124.140501,Bezerra_2025}, the centroid and separation of two incoherent point sources \cite{PhysRevA.96.062107,PhysRevA.95.063847,PhysRevA.98.012103,PhysRevA.99.022116}, all parameters in a Lie group \cite{y9wq-hhg2,hou2021zero,537s-r3t8}, to name a few.
The former may be used for scalable and programmable linear network tasks, while the latter is capable of analyzing some practical sensing scenarios.

In this work, we address the simultaneous estimation of a phase shift and a transmittance parameter in a two-channel optical network.
The network is composed of a beam splitter and a phase shifter, which is the simplest example in multi-mode distributed quantum sensing \cite{Gramegna_2021,PhysRevA.104.062603}.
This combination can also be regarded as the kernel of a general linear interferometer, as it can entirely determine the quantum Fisher information (QFI) of the interferometer \cite{PhysRevA.106.023716,PhysRevA.107.043704,PhysRevA.105.012604}.
We deploy a Schr\"odinger's cat state along with a coherent state as the probe states.
At the readout stage, a two-mode projective measurement is performed onto the output states.
The parameter sensitivities are discussed, and the effects of photon loss are analyzed.
Our findings are expected to enrich scenarios for multi-parameter estimation, thus facilitating the development of quantum precision measurements.

The rest of this paper is organized as follows.
Section \ref{II} introduces the two-channel optical network and analyzes the sensitivity limit through the QFI matrix.
In Sec. \ref{III}, we show the specific measurement and calculates the sensitivities of two parameters via the CFI analysis.
The effects of photon loss on parameter sensitivities are studied in Sec. \ref{IV}.
Finally, we summarize the main results in this work in Sec. \ref{V}.

\section{Fundamental protocol and sensitivity limit}
\label{II}

\begin{figure}[htbp]
	\centering
	\includegraphics[width=0.25\textwidth]{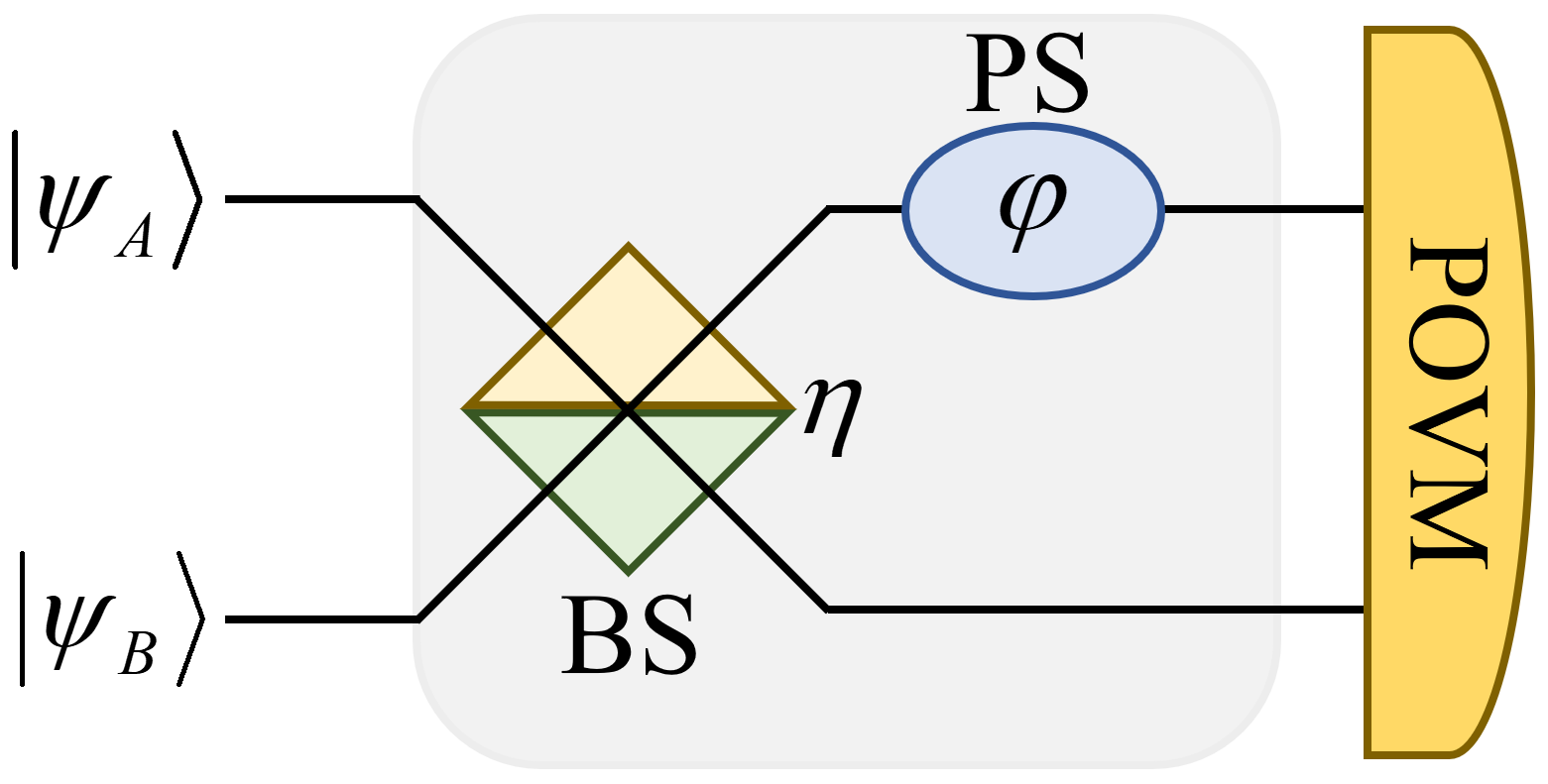}
	\caption{The illustration of the estimation protocol using a two-channel optical network, BS, beam splitter; PS, phase shifter; POVM, positive operator valued measure.}
	\label{system}
\end{figure}

Figure \ref{system} shows the illustration of two-parameter estimation protocol based on the two-channel optical network.
The transmittance $\eta$ of the beam splitter and the phase difference $\varphi$ between the two modes are parameters we would like to estimate.
A Schr\"odinger's cat state along with a coherent state are utilized as the probe states.
To be specific, the two-mode probe states can be written as $\left| {{\psi _{{\rm{in}}}}} \right\rangle  = \left| {{\psi _A}} \right\rangle  \otimes \left| {{\psi _B}} \right\rangle $ with
 $\left| {{\psi _A}} \right\rangle  = {N_\alpha }\left( {\left| {{\alpha  \mathord{\left/
 {\vphantom {\alpha  {\sqrt 2 }}} \right.
 \kern-\nulldelimiterspace} {\sqrt 2 }}} \right\rangle  + \left| { - {\alpha  \mathord{\left/
 {\vphantom {\alpha  {\sqrt 2 }}} \right.
 \kern-\nulldelimiterspace} {\sqrt 2 }}} \right\rangle } \right)$ and $\left| {{\psi _B}} \right\rangle  = \left| {{\alpha  \mathord{\left/
 {\vphantom {\alpha  {\sqrt 2 }}} \right.
 \kern-\nulldelimiterspace} {\sqrt 2 }}} \right\rangle $, where $\alpha$ is amplitude parameter and ${N_\alpha } = {[ {2( {1 + {e^{ - {\alpha ^2}}}} )} ]^{ - 1/2}}$ stands for normalization factor. 
Related to this, the average photon number in our protocol is given by 
\begin{align}
N = \langle {{{\hat a}^\dag }\hat a + {{\hat b}^\dag }\hat b} \rangle  = 2{\alpha ^2}N_\alpha ^2,
\label{e1}
\end{align}
where $\hat a^{\dagger}$ ($\hat b^{\dagger}$) and $\hat a$ ($\hat b$) are creation and annihilation operators of mode $A$ ($B$), respectively.

In the two-channel optical network, probe states of two modes are combined by a beam splitter with 
transmittance $\eta$.
Upon leaving the beam splitter, the states evolve into
\begin{align}
{\hat B_\eta }\left| {{\psi _{{\rm{in}}}}} \right\rangle  = {N_\alpha }\left( {\left| {{k_1}\alpha } \right\rangle \left| {{k_2}\alpha } \right\rangle  + \left| {-{k_2}\alpha } \right\rangle \left| {{k_1}\alpha } \right\rangle } \right)
\label{e2}
\end{align}
with ${k_1} = \sqrt {{\eta  \mathord{\left/
 {\vphantom {\eta  2}} \right.
 \kern-\nulldelimiterspace} 2}} + \sqrt {{{(1 - \eta )} \mathord{\left/
 {\vphantom {{(1 - \eta )} 2}} \right.
 \kern-\nulldelimiterspace} 2}} $ and ${k_2} = \sqrt {{\eta  \mathord{\left/
 {\vphantom {\eta  2}} \right.
 \kern-\nulldelimiterspace} 2}}  - \sqrt {{{(1 - \eta )} \mathord{\left/
 {\vphantom {{(1 - \eta )} 2}} \right.
 \kern-\nulldelimiterspace} 2}} $.
The specific operator form of the beam splitter is ${\hat B_\eta } = \exp [ {\arccos \sqrt \eta  ( {{{\hat a}^\dag }\hat b - \hat a{{\hat b}^\dag }} )} ]$.
Then a phase difference $\varphi$ is imprinted on the states in Eq. (\ref{e2}), and the parameterized states, the output states of the network, can be expressed as
\begin{align}
\nonumber \left| {{\psi _p }} \right\rangle  &= {\hat U_\varphi }{\hat B_\eta }\left| {{\psi _{{\rm{in}}}}} \right\rangle  \\
&= {N_\alpha }\left( {\left| {{k_1}\alpha {e^{{\rm i}\varphi }}} \right\rangle \left| {{k_2}\alpha } \right\rangle  + \left| { - {k_2}\alpha {e^{{\rm i}\varphi }}} \right\rangle \left| {{k_1}\alpha } \right\rangle } \right).
\label{}
\end{align}

Based on the parameterized states, one can perform the QFI analysis.
For such two-parameter problem, the sensitivity limits are determined by a two-by-two QFI matrix of which elements are as follows
\begin{align}
{Q_{mn}} = 4{\mathop{\rm Re}\nolimits} \left[ {\frac{{\partial \left\langle {{\psi _p }} \right|}}{{\partial {p _m}}}\frac{{\partial \left| {{\psi _p}} \right\rangle }}{{\partial {p _n}}} - \frac{{\partial \left\langle {{\psi _p }} \right|}}{{\partial {p _m}}}\left| {{\psi _p }} \right\rangle \left\langle {{\psi _p }} \right|\frac{{\partial \left| {{\psi _p }} \right\rangle }}{{\partial {p _n}}}} \right],
\label{}
\end{align}
where $m,n \in \left\{ {1,2} \right\}$ with
${p _1} = \varphi$ and ${p _2} = \eta $. 
A straightforward calculation shows that non-zero elements are
\begin{align}
\nonumber {Q_{11}} =& \frac{{{\alpha ^2}}}{{1 + {e^{ - {\alpha ^2}}}}}\{ {\left[ {1 + 4\eta \left( {1 - \eta } \right)} \right]{\alpha ^2} - 2\left( {2\eta  - 1} \right){e^{ - {\alpha ^2}}}}  \\
\nonumber &+ \left[ {1 - 4\eta \left( {1 - \eta } \right)} \right]{\alpha ^2}{e^{ - {\alpha ^2}}} + 2 \} - \frac{{{\alpha ^4}}}{{{{( {1 + {e^{ - {\alpha ^2}}}} )}^2}}}\\
 &\times[ {1 - \left( {2\eta  - 1} \right){e^{ - {\alpha ^2}}}} ]^2,
\label{e5}
\end{align}
\begin{align}
{Q_{22}} = \frac{{{\alpha ^2} - {\alpha ^4}{e^{ - {\alpha ^2}}}}}{{\eta \left( {1 - \eta } \right)\left( {1 + {e^{ - {\alpha ^2}}}} \right)}},
\label{e6}
\end{align}
and ${Q_{12}} = {Q_{12}} = 0$.
The diagonal QFI matrix indicates compatibility between the two parameters, meaning that simultaneously optimal two-parameter estimation is attainable. 
It is noteworthy that two matrix elements merely depend on the transmittance parameter. 
This originates from different properties of the encoding processes of the two parameters.
Specifically, the generator ${\rm i}( {\partial _\varphi ^{}\hat U_\varphi ^\dag } )\hat U_\varphi ^{}$ is independent of the phase shift $\varphi$, while the generator ${\rm i}( {\partial _\eta ^{}\hat B_\eta ^\dag } )\hat B_\eta ^{}$ is associated with the transmittance parameter $\eta$.
Using the QFI matrix, we can obtain the sensitivity limits of two parameters 
\begin{align}
\Delta \varphi_{\rm min}  &= \frac{1}{{\sqrt {Q_{22}^{ - 1}} }} = \frac{1}{{\sqrt {{Q_{11}}}}}, \\
\Delta \eta_{\rm min}   &= \frac{1}{{\sqrt {Q_{11}^{ - 1}} }} = \frac{1}{{\sqrt {{Q_{22}}} }}. 
\label{}
\end{align}

\begin{figure}[htbp]
	\centering
	\includegraphics[width=0.48\textwidth]{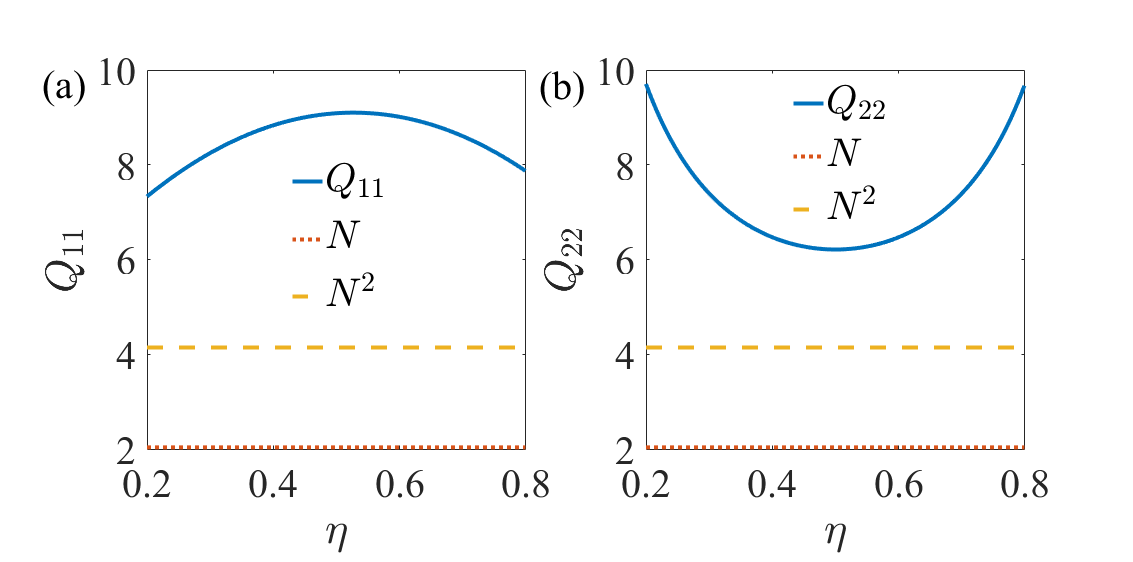}
	\caption{The elements of the QFI matrix against phase shift $\varphi$ and transmittance parameter $\eta$ with amplitude parameter $\alpha = 1$, (a) $Q_{11}$; (b) $Q_{22}$. }
	\label{f2}
\end{figure}

Figure \ref{f2} shows the diagonal elements $Q_{11}$ and $Q_{22}$ of the QFI matrix.
Both $Q_{11}$ and $Q_{22}$ are higher than $N^2$, meaning that the sensitivity limits are superior to the Heisenberg limit for low-photon region.
Meanwhile, $Q_{11}$ is asymmetric about $\eta=0.5$ and $Q_{22}$ is symmetric about $\eta=1/2$, which can be seen from Eqs. (\ref{e5}) and (\ref{e6}).
With the increase of amplitude parameter $\alpha$, the average photon number is $N \approx {\alpha^2}$, and the diagonal elements are recast as
\begin{align}
{Q_{11}} \xrightarrow{\alpha \gg 0}{4\eta \left( {1 - \eta } \right){\alpha ^4}},
\label{}
\end{align}
\begin{align}
{Q_{22}} \xrightarrow{\alpha \gg 0} \frac{1 }{{\eta \left( {1 - \eta } \right)}}{\alpha ^2}.
\label{e10}
\end{align}
The result given by Eq. (\ref{e10}) is consistent with that in Ref. \cite{PhysRevLett.121.230801}. 
For high-photon region, the sensitivity limits of phase estimation and transmittance estimation are the Heisenberg limit and the sub-shot-noise limit, respectively.
Especially, the phase estimation and transmittance estimation reach the optimal and the worst sensitivity limits with a balanced beam splitter ($\eta = 1/2$).

\section{Sensitivities with projective measurement}
\label{III}

In order to obtain the parameter information, one needs to implement a specific positive operator valued measure (POVM) to the output states.
Here we consider a two-mode projective operator which has the following representation in twin Fock basis
\begin{align}
{\hat \mu _{AB}} = \sum\limits_{m = 0}^\infty  {\sum\limits_{n = 0}^\infty  {\left| {m} \right\rangle \left| {n} \right\rangle \left\langle {n} \right| \left\langle {m} \right|} } .
\label{}
\end{align}
This POVM can be realized by a balanced beam splitter followed by parity detection \cite{PhysRevA.99.022106}, i.e., ${\hat \mu _{AB}} = {{\hat B}^{\dagger}_{1/2}}{{\hat \Pi }^{}_A}{{\hat B}^{}_{1/2}}$.
The specific operator form of a balanced beam splitter is ${\hat B_{1/2} } = \exp [ {\pi  ( {{{\hat a}^\dag }\hat b - \hat a{{\hat b}^\dag }} )/4} ]$, and ${\hat \Pi _A} = \exp ( {{\rm i}\pi {{\hat a}^\dag }\hat a} )$ is parity operator.

The parameterized states after a balanced beam splitter are found to be
\begin{align}
\nonumber \left| {{\psi }} \right\rangle  =& {{\hat B}_{1/2}}\left| {{\psi _p }} \right\rangle \\
\nonumber =& {N_\alpha } \left| {\frac{\alpha }{{\sqrt 2 }}\left( {{k_1}{e^{{\rm i}\varphi }} - {k_2}} \right)} \right\rangle \left| {\frac{\alpha }{{\sqrt 2 }}\left( {{k_1}{e^{{\rm i}\varphi }} + {k_2}} \right)} \right\rangle  \\
 &+ {N_\alpha }\left| { - \frac{\alpha }{{\sqrt 2 }}\left( {{k_1} + {k_2}{e^{{\rm i}\varphi }}} \right)} \right\rangle \left| {\frac{{ \alpha }}{{\sqrt 2 }}\left( {{k_1} - {k_2}{e^{{\rm i}\varphi }}} \right)} \right\rangle.  
\label{}
\end{align}
Based on the above results, we can obtain the readout signal, i.e., the expectation value of projective operator
\begin{align}
\nonumber \langle {\hat \mu _{AB}} \rangle  =& {N^2_\alpha }( \exp\{ - \left[ {1 - \left( {1 - 2\eta } \right)\cos \varphi } \right]{\alpha ^2}\} \\
\nonumber&+ \exp\{ - \left[ {1 + \left( {1 - 2\eta } \right)\cos \varphi } \right]{\alpha ^2}\} \\
\nonumber&+ \exp\{ - {\alpha ^2} - {\alpha ^2}[ {{\rm i}\sin \varphi  - 2\sqrt {\eta \left( {1 - \eta } \right)} \cos \varphi } ]\}\\
&+ \exp\{ - {\alpha ^2} + {\alpha ^2}[ {{\rm i}\sin \varphi  + 2\sqrt {\eta \left( {1 - \eta } \right)} \cos \varphi } ]\} ),
	\label{}
\end{align}
where the expectation value are taken over by state $\left| {{\psi }} \right\rangle$.
In the calculation process, we utilize the inner product of coherent states 
\begin{align}
\left\langle {{{\beta _1}}}
 {\left | {\vphantom {{{\beta _1}} {{\beta _2}}}}
 \right. \kern-\nulldelimiterspace}
 {{{\beta _2}}} \right\rangle  = \exp \left( { - \frac{1}{2}{{\left| {{\beta _1}} \right|}^2} - \frac{1}{2}{{\left| {{\beta _2}} \right|}^2} + \beta _1^ * {\beta _2}} \right)
\label{e7}
\end{align}
and
\begin{align}
\left\langle {{\beta _1}} \right|{\hat \Pi _A}\left| {{\beta _2}} \right\rangle  = \exp \left( { - \frac{1}{2}{{\left| {{\beta _1}} \right|}^2} - \frac{1}{2}{{\left| {{\beta _2}} \right|}^2} - \beta _1^ * {\beta _2}} \right).
\label{e9}
\end{align}

\begin{figure}[htbp]
	\centering
	\includegraphics[width=0.48\textwidth]{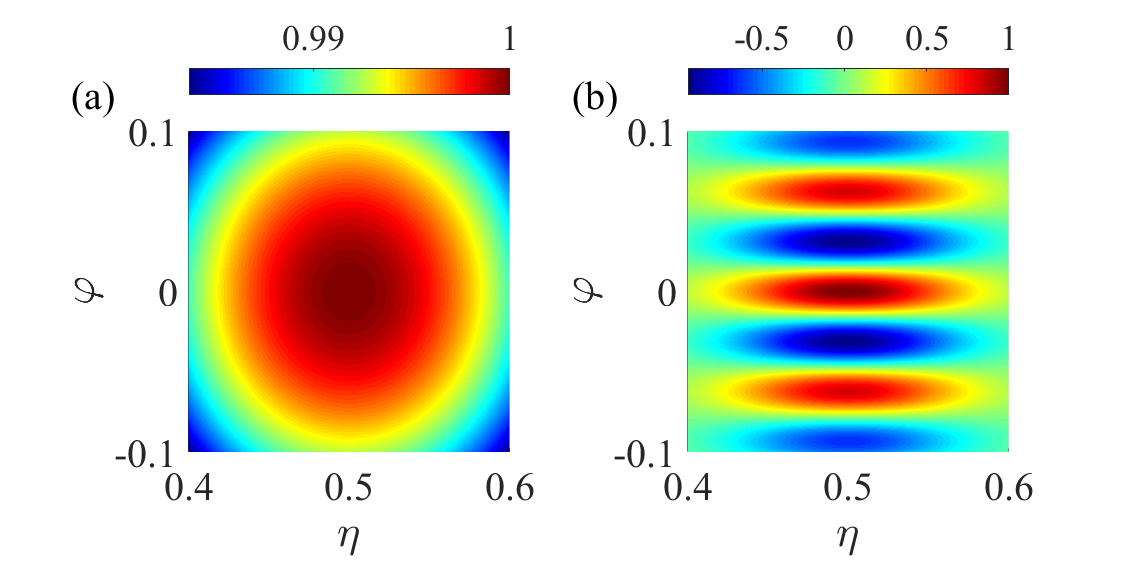}
	\caption{The readout signal against phase shift $\varphi$ and transmittance $\eta$, (a) $\alpha = 1$; (b) $\alpha = 10$. }
	\label{f3}
\end{figure}

In Fig. \ref{f3}, we give the dependence of the readout signal on the phase shift and transmittance parameter. 
For low-photon region, the readout signal varies slowly with the two estimated parameters. 
As the number of photons increases, the oscillation of the readout signal increases significantly. 
This demonstrates that the resolution of our protocol is improved. 
When compared with transmittance parameter, the readout signal is more sensitive to the phase shift.

In general, one needs a photon-number-resolving detector to perform parity detection. 
The detection event is recorded as +1 ($-1$) if the detector is clicked by an even (odd) number of photons.
As a result, the probability of detecting an even or odd number of photons can be written as
\begin{align}
{P_ + } &= \frac{1}{2}\left( {1 + \langle {\hat \mu _{AB}} \rangle } \right),\\
{P_ - } &= \frac{1}{2}\left( {1 - \langle {\hat \mu _{AB}} \rangle } \right).
\label{}
\end{align}

\begin{figure*}[htbp]
	\centering
	\includegraphics[width=0.85\textwidth]{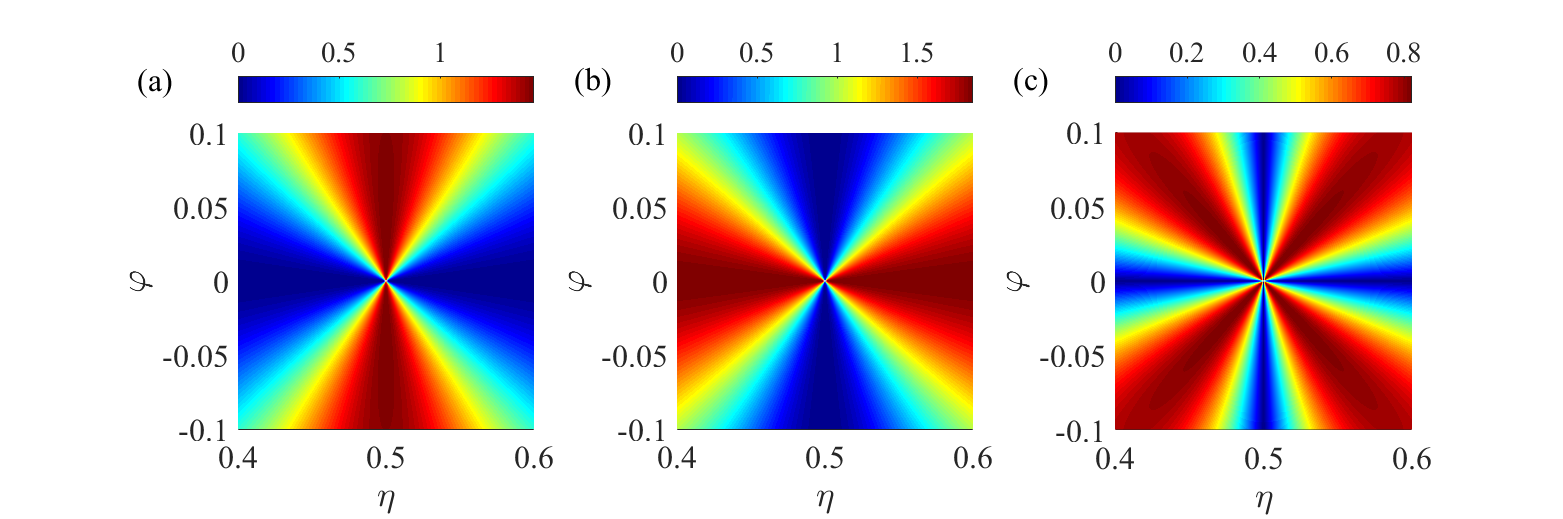}\\
    \includegraphics[width=0.85\textwidth]{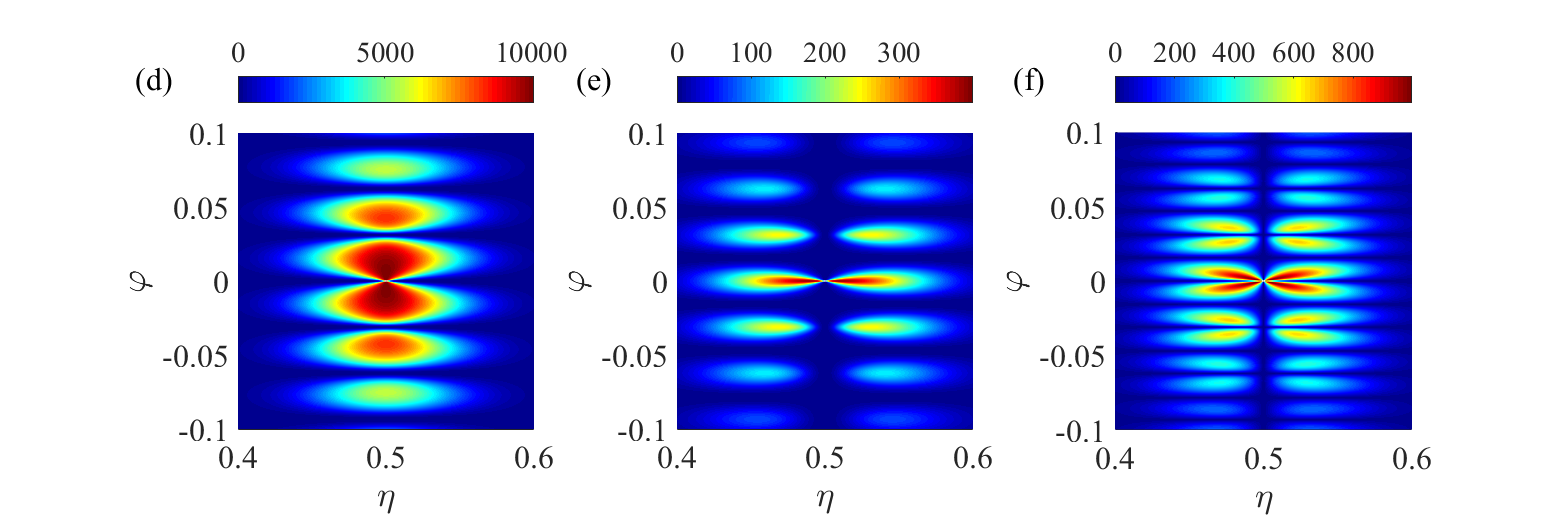}
	\caption{The elements of the CFI matrix against phase shift $\varphi$ and transmittance parameter $\eta$, (a) $C_{11}$ with $\alpha = 1$; (b) $C_{22}$ with $\alpha = 1$; (c) $|C_{12}|$ with $\alpha = 1$; (d) $C_{11}$ with $\alpha = 10$; (e) $C_{22}$ with $\alpha = 10$; (f) $|C_{12}|$ with $\alpha = 10$.}
	\label{f4}
\end{figure*}

Using these two probabilities, we can further calculate the elements of CFI matrix in terms of
\begin{align}
{C_{mn}} = \frac{1}{{{P_ + }}}\frac{{\partial {P_ + }}}{{\partial {p _m}}}\frac{{\partial {P_ + }}}{{\partial {p _n}}} + \frac{1}{{{P_ - }}}\frac{{\partial {P_ - }}}{{\partial {p _m}}}\frac{{\partial {P_ - }}}{{\partial {p _n}}},
\label{}
\end{align}
where $m,n \in \left\{ {1,2} \right\}$ with
${p _1} = \varphi$ and ${p _2} = \eta $.
The corresponding sensitivities are given by
\begin{align}
\Delta \varphi &= \left[{{C}_{\varphi \varphi }} - \frac{{{{C}_{\varphi \eta }}{{C}_{\eta \varphi }}}}{{{{C}_{\eta \eta }}}}\right]^{-1/2},  
\label{}\\
\Delta \eta &= \left[{{C}_{\eta \eta }} - \frac{{{{C}_{\varphi \eta }}{{C}_{\eta \varphi }}}}{{{{C}_{\varphi \varphi }}}}\right]^{-1/2}.
\label{}
\end{align}
The off-diagonal elements in the CFI matrix are generally non-zero, which are different from their counterparts in the QFI matrix.
This fact indicates that two parameters are an interactive relationship with respect to the used POVM.
In particular, only when $\eta  = {1}/{2}$ and $\varphi  = 0$ are satisfied can we obtain ${C_{12}} = {C_{21}} = 0$.

To intuitively observe the CFI matrix, in Fig. \ref{f4} we plot matrix elements as a function of the phase shift and transmittance parameter. 
The results indicates that diagonal elements reach the maximal values and the off-diagonal elements vanish with $\eta  = {1}/{2}$ and $\varphi  = 0$.
As the phase shift $\varphi$ deviates from 0, the element ${C_{11}}$ exhibits an damped decay.
The element ${C_{22}}$ shows similar decay when the transmittance parameter $\eta$ deviated from $1/2$.
These conditions reveal the optimal working points in our protocol.

Now we compare the sensitivities and the corresponding limits to evaluate the optimality of the used POVM. 
Here we introduce achievement rate which is defined as
\begin{align}
{\cal A}_{\varphi} = \frac{\Delta \varphi_{\rm min}}{{\left. {\Delta \varphi } \right|_{\varphi  = 0, \, \eta  = 1/2}}},\quad {\cal A}_{\eta} = \frac{\Delta \eta_{\rm min}}{{\left. {\Delta \eta } \right|_{\varphi  = 0, \, \eta  = 1/2}}},
\label{}
\end{align}
where we have $0 \le {\cal A}_{\varphi} \le 1$ and $0 \le {\cal A}_{\eta} \le 1$.

\begin{figure}[htbp]
	\centering
	\includegraphics[width=0.48\textwidth]{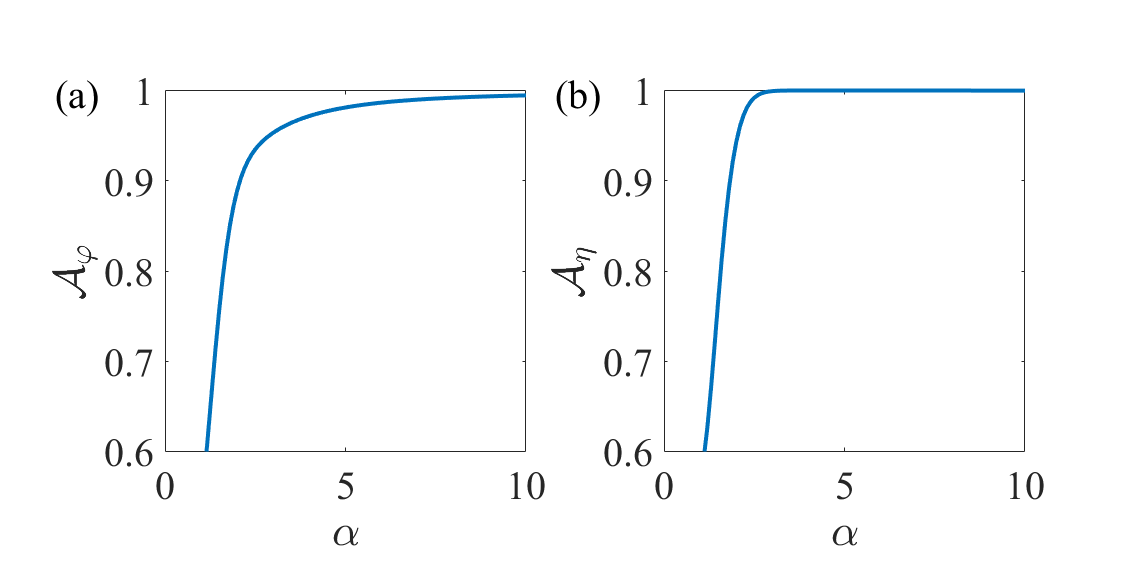}
	\caption{The achievement rate against amplitude parameter $\alpha$, (a) ${\cal A}_{\varphi}$; (b) ${\cal A}_{\eta}$. }
	\label{f5}
\end{figure}

Figure \ref{f5} illustrates the relationship between achievement rate and amplitude parameter. 
The achievement rate of transmittance parameter is superior to that of phase shift by comparison.
In low-photon region, these two achievement rates can reach approximately 60\%. 
With the increase of the average photon number, both of them gradually approach 100\%.
This proves, in high-photon region, the approximate optimality of the used POVM for our two-parameter estimation problem.

\begin{figure}[htbp]
	\centering
	\includegraphics[width=0.32\textwidth]{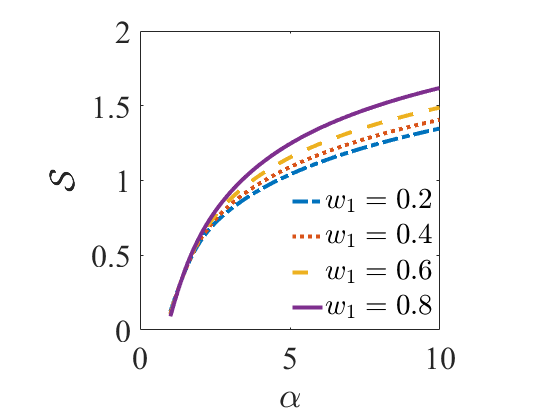}
	\caption{The logarithmic sensitivity parameter $\cal S$ against amplitude parameter $\alpha$ with different weights. }
	\label{f6}
\end{figure}

In some practical scenarios, there is a trade-off between the two parameters; more attention is focused on one of the parameters to be estimated.
A feasible approach in these scenarios is to consider the figure of merit as \cite{PhysRevLett.125.020501}
\begin{align}
\Delta^2 _{\rm w} = {{w_1}\Delta^2 \varphi_{\rm min} + {w_2}\Delta^2 \eta_{\rm min}},
\label{}
\end{align}
which is weighted sensitivity with respect to these two parameters.
Note that weights of two parameters, $w_1$ and $w_2$,
take value in the interval $[0, 1]$, and we have $w_1 + w_2 = 1$.
To quantitatively evaluate the weighted sensitivity, we can further calculate logarithmic sensitivity parameter
\begin{align}
{\cal S} =  - {\log _N}{\Delta^2 _{\rm{w}}}.
\label{}
\end{align}
When the weighted sensitivity reaches the shot-noise limit and Heisenberg limit, this parameter sit at ${\cal S} = 1$ and ${\cal S} = 2$.
For our two-parameter estimation problem, as shown in Fig. \ref{f6}, it turns out that the parameter satisfies ${\cal S} > 1$ for high-photon region.
This demonstrates the sensitivity advantage originating from quantum enhancement.

\section{Effects of photon loss on sensitivity}
\label{IV}

In the previous section, we address the parameter sensitivities in a lossless scenario. 
For practical quantum sensing tasks, photon loss is an unavoidable challenge and a major constraint on the performance. 
It disrupts the coherence, squeezing and entanglement in quantum states, causing the enhanced advantage from quantum resources to drastically decay.
In this section, we pay our attention to the parameter sensitivities in the presence of photon loss.

In terms of the theory of quantum open systems, photon loss occurs in a lossy channel.
This process can be described theoretically by a Kraus operator and modeled experimental by a virtual beam splitter.
Generally, a pure state upon leaving a lossy channel becomes a mixed state, which corresponds to a decrease in coherence. 
In particular, a coherent state experiencing a lossy channel remains itself except decrease in amplitude.
This is advantageous for our protocol, in that the parameterized states are entangled coherent states.
For the sake of simplicity, we assume that the lossy rates of the two modes in the network are the same, and the total lossy rate of the entire network is $L$.
Related to this, the sensitivity of a lossy $N$-photon protocol is equal to the that of a protocol fed by $LN$ photons without loss.

\begin{figure}[htbp]
	\centering
	\includegraphics[width=0.42\textwidth]{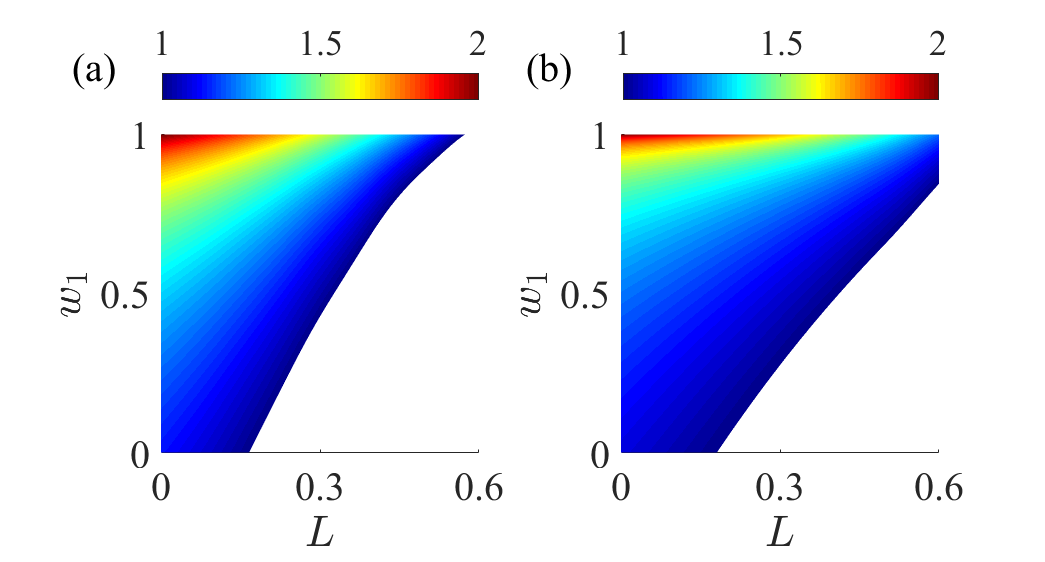}
	\caption{The logarithmic sensitivity parameter $\cal S$ against lossy rate $L$ and weight $w_1$, (a) $\alpha = 5$; (b) $\alpha = 10$. The logarithmic sensitivity parameter corresponding to the blank area is lower than 1.}
	\label{f7}
\end{figure}

Based on the above results, we can address the parameter sensitivity in the presence of photon loss.
Figure \ref{f7} gives the variation of the logarithmic sensitivity parameter with lossy rate and weight.
An intuitive phenomenon is that, with the increase of weight $w_1$, the tolerance to photon loss is significantly improved.
This is due to the fact that the sensitivity of phase estimation surpasses that of transmittance estimation. 
Concurrently, the entanglement between two modes increases with increasing average photon number and shows advantages in resisting photon loss.
As a consequence, the loss-tolerance can be also improved by increasing the average photon number.
The result indicates that our protocol is of excellent loss-tolerance.
In the most challenging scenarios ($w_1 \rightarrow 0$), the sensitivity can surpass the shot-noise limit at a lossy rate of about 20\%.
For other scenarios, the maximum tolerable lossy rate corresponding to the sub-shot-noise-limited sensitivity will further increase.

\section{Conclusion}
\label{V}
In summary, we investigate the problem of simultaneous phase and transmittance estimation in a two-channel optical network.
A Schr\"odinger’s cat state along with a coherent state are injected into the network, and projective measurement is selected as the POVM for readout.
The readout signal is super-resolving with multiple oscillations.
For a lossless high-photon scenario, the CFI analysis shows that the sensitivity of phase estimation and that of transmittance estimation can reach the Heisenberg limit and the sub-shot-noise limit, respectively.
These sensitivities are saturated at the sensitivity limits given by the QFI matrix.
In addition, we study the effects of photon loss and introduce weighted sensitivity to comprehensively evaluate the performance.
The results indicate that our protocol is loss-tolerant and sensitivity beyond the shot-noise limit is attainable at a lossy rate of 20\%.
Our findings enrich the tasks of quantum information processing.
Future work will likely focus on the application of quantum multi-parameter estimation in quantum optical neural networks.

\section*{Acknowledgment} 
This work was supported by the National Natural Science Foundation of China (U24A2017), Program of Zhongwu Young Innovative Talents of Jiangsu University of Technology (20230013), and Qing Lan Project of Higher Education of Jiangsu Province of China.


%

\end{document}